# Hybrid Intelligent Routing with Optimized Learning (HIROL) for Adaptive Routing Topology management in FANETs

Ch. Naveen Kumar Reddy[1], Dr. M. Anusha[2]

[1] Research Scholar, Dept. of CSE, Koneru Lakshmaiah Education Foundation, AP, India.
naveenkumarreddy.ch@gmail.com
[2] Assoc.Professor, Dept. of CSE, Koneru Lakshmaiah Education Foundation, AP, India.
anushaaa9@kluniversity.in

**Abstract**

Enhancing the routing efficacy of Flying AdHoc Networks (FANETs), a network of numerous Unmanned Aerial Vehicles (UAVs), in which various challenges may arise as a result of the varied mobility, speed, direction, and rapid topology changes. Given the special features of UAVs, in particular their fast mobility, frequent topology changes, and 3D space movements, it is difficult to transport them through a FANET. The suggested study presents a complete hybrid model: HIROL (Hybrid Intelligent Routing with Optimized Learning) that integrates the ABC (Artificial Bee Colony) algorithm, DSR (Dynamic Source Routing) by incorporating Optimized Link State Routing (OLSR) and ANNs (Artificial Neural Networks) to optimize the routing process. The HIROL optimizes link management by ABC optimization algorithm and reliably analyses link status using characteristics from OLSR and DSR; at the same time, an ANN-based technique successfully classifies connection state. In order to provide optimal route design and maintenance, HIROL dynamically migrates between OLSR and DSR approaches according to the network topology conditions. After running thorough tests in Network Simulator 2 (NS-2), when compared to more conventional DSR and OLSR models, the hybrid model HIROL performs far better in simulations and tests. An increase in throughput (3.5 Mbps vs. 3.2-3.4 Mbps), a decrease in communication overhead (15% vs. 18-20%), and an improvement in Packet Delivery Ratio (97.5% vs. 94-95.5%). These results demonstrate that the suggested HIROL model improves FANET routing performance in different types of networks.

## 1. Introduction

Recently, there has been a lot of interest in using UAVs for communication due to how quickly technology is developing. Today's electronics, computers, gadgets, and media are commercial products, and as a result, they have evolved into unmanned machines that include balloons, drones, and tiny aircraft [1]. Drone communications, also referred to as Flying Ad hoc Networks (FANETs), have applications in several domains, such as advanced communications, agriculture, disaster management, and surveillance. The distinct network structure, the unpredictable and sporadic motions of UAVs, and the restricted resources of these vehicles provide hurdles to the effective operation of data packets in FANET. Conventional routing protocols intended for ad hoc networks on ground are frequently inadequate for handling the particular resources and constraints associated with FANET challenges [2].

FANET routing protocol capabilities can now be enhanced through the integration of effective routing algorithms with Artificial Intelligence (AI) methods, especially Neural Networks. In order to achieve the best possible routing in FANETs, this work introduces a novel hybrid approach called HIROL, which blends dynamic communication approaches with neural network optimization. Integrating these two routing techniques aims at maximizing their respective benefits, reduce the limitations of traditional routing methodologies, and enhance the effectiveness and simplicity of routing in dynamic FANET systems.

The key issues with FANET routing are outlined in this prologue, which also addresses the objectives and guidelines of the competition model and looks at the constraints of the available methods for routing in addition to opportunities for neural networks to enhance decisions regarding routing.

### 1.1. Difficulties in FANET Routing

The wide range of drones and network circumstances renders routing in FANETs a difficult responsibility. The following problems exist in FANET.

- Unlike traditional networks, FANETs require high speed and a way to adapt to frequent dynamic changes while ensuring efficient data transmission. Most of the traditional protocols do not adequately meet FANETs' requirements. Reducing signaling overhead and ensuring consistent end-to-end data transmission are made possible by routing protocols [3]
- Energy, computing power and communication are very limited in these areas. Efficient use of energy is important to extend the life of drones and maintain uninterrupted connectivity.[4]
- Unmanned Aerial Vehicles (UAVs) also have different communication systems depending on their intended use, requiring instant information transfer, advancing important information and having the ability to handle many things [5].

The different routing protocols of FANET are represented in Fig. 1

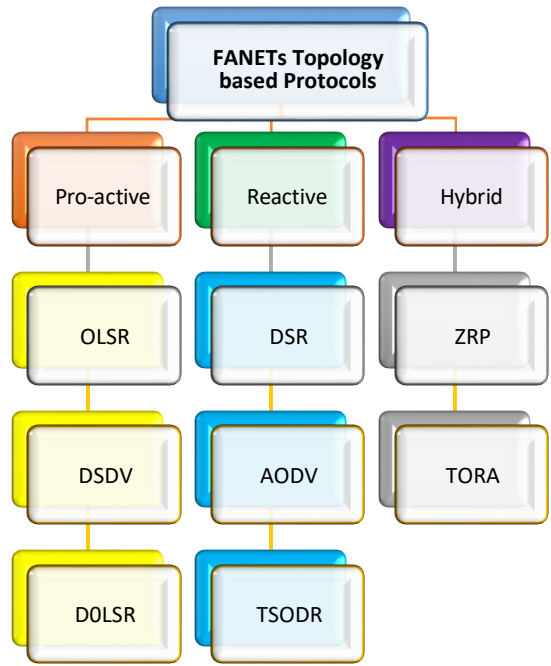

Fig.1: Routing protocols of FANET

**1.2. Constraints of Current Routing Protocols:**

The routing algorithm D-OLSR is used to choose the Multi-Point Relay (MPR) and can decrease the number of relays with directional aerials as shown in Figire 2. The D-OLSR seeks to regulate quantity of MPR in given network so as to decrease the network overhead and delay [6]. Implementing directional antennas can create more complexity to the design of network and configuration.

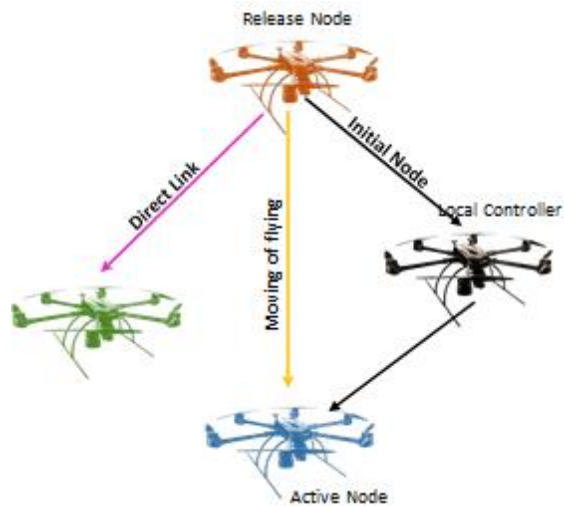

Fig.2: Process of DOLSR

Fig.3 shows how the Destination-Sequenced-Distance-Vector (DSDV) relies on a sequence number for each node in a routing table to ensure that the protocol does not loop. DSDV routing, pose challenges when operating on FANETs. These methods rely on continuous sharing of information or detection methods, which can lead to further interference and delays in FANET systems [7]. They will also be unable to adapt to UAV behaviour or make better decisions for power management and network circumstances.

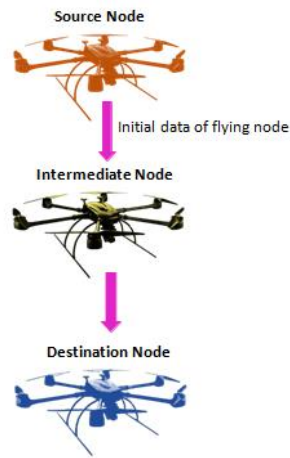

Fig.3: Process of DSDV

Existing routing protocols for FANETs generally do not include mechanisms for determining various Quality of Service (QoS) requirements. These requirements are crucial for applications such as real-time multimedia streaming or the main purpose of data transfer [8]. In the absence of good QoS awareness of traffic, FANETs will face difficulties in achieving performance targets and maintaining reliable communication in different situations.

Adaptive routing protocols like DSR enable self-organization and self-configuration of the network without requiring any infrastructure. Owing to DSR's reactive structure, a discovery process is only initiated in response to a communication need. To maintain any path failures, a route maintenance mechanism is also implemented. In order to initiate route recovery, DSR was utilized by the source node to notify neighboring nodes. Because every exchanged packet must contain every address of every transited node, it is insufficient for both huge networks and topologically extremely dynamic networks [9].

The AODV (Adhoc On-demand Distance Vector) method exists especially for routing table maintenance and the source node keeps the next information of the network as shown in Fig. 4. It acquires recurring updates from DSDV as well as hop-to-hop routing by means of DSR. Due to its reactive design, AODV only locates a path when it is required and ignores paths that lead to locations that are not engaged in active communication. [10]

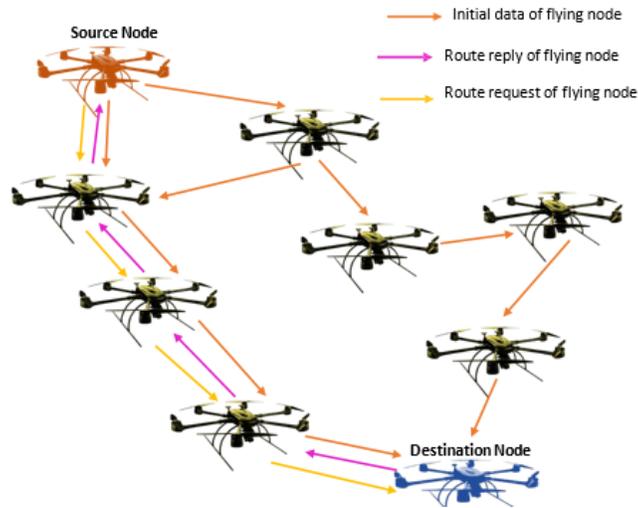

Fig. 4: Process of AODV

Artificial Neural Networks (ANNs) have shown great potential in capturing complex patterns, predicting outcomes, and improving decision-making processes in many fields. Neural networks have the ability to improve FANET routing by increasing routing efficiency, responding to changing network conditions, and optimizing services Using historical information and the current state of the network, neural networks can learn to predict patterns such as changes in node location, traffic patterns, and network quality. FANET can benefit through decision making, advancements in technology, and network dynamics adaptability through the integration of the neural network model into a formal framework [11]. Neural networks are additionally utilized for QoS-aware routing, intelligent load balancing, and optimization of routes, all of which enhance network efficiency and the user experience.

The primary objective of the proposed system is to create an innovative hybrid FANET routing strategy which combines efficient routing methods with artificial neural networks. The primary goals of the proposed hybrid approach are outlined below:

- To provide effectiveness in routing methods which have been tailored to the resource-limited and dynamic nature of FANETs, taking into account parameters such as network behavior, node movement, and consumption of energy.
- To build models of neural networks with the capacity to acquire and foresee network actions, particularly patterns of traffic, node motions, and variations in link quality.
- To produce intelligent, adaptive, and QoS-aware routing decisions in FANETs through the integration of neural network-based optimization approaches with routing techniques.
- To determine the hybrid model's performance through contrasting it to the present routing protocols and evaluating crucial parameters including throughput, latency, energy efficiency, and scalability.

To assess whether the hybrid model is suitable for use in different FANET scenarios, including applications in industry, aerial networks for communication, emergencies, and monitoring operations.

**1.3. Paper Structure**

A comprehensive overview of previous studies on FANET routing, covering the newest protocols, optimization methods, and AI-based strategies, is presented in Section 1. The literature survey and related works are discussed in section 2.Section 3 outlines the proposed work of the hybrid model, explaining how efficient routing algorithms are integrated with neural networks. Section 4 examines the findings, explores the ramifications of the results, and emphasizes the contributions of the proposed model to the field of FANET routing. Section 5 delineates the prospective avenues for further research, and final observations.

To summarize, combining effective routing algorithms with neural networks offers a hopeful method for tackling the obstacles of optimized routing approach in FANETs. The proposed hybrid model combines adaptive learning capabilities of neural networks with the domain-specific optimizations of routing algorithms. Its goal is to improve the efficiency, resilience, and scalability of FANETs, enabling advanced applications in aerial communication and unmanned aerial systems.

**2. Literature survey and related works**

In this study, we categories and analyses the existing communication algorithms and routing protocols employed in FANETs and similar environments based on different approaches, such as  proactive, reactive, hybrid, position-based and cluster methodologies.

**2.1. Proactive Routing (Table Driven):**

Drones employ proactive routing, a strategy in which they consistently update and exchange their navigation information. Every node within the network shares its routes with every other node.. To minimize waiting periods, the network may select the routing path between each pair of drones. However, there are clear disadvantages as well. One notable downside is the substantial rise in communication overhead caused by the vast quantity of control packets required to maintain information related to routing. In addition, high-mobility networks are not suitable for proactive routing strategies. For their work on monitoring the traffic, authors in [12] relied on OLSR approach. It is concluded from their simulation results that the OLSR's considerable overhead makes it suboptimal for high dynamic, low density FANETs. The flooding approach of OLSR can increase the routing table overhead in large scale networks

The authors of [13] proposed a speed attentive Predictive-Optimized Link State Routing methodology (P-OLSR) to improve FANET routing operations. This methodology uses GPS data to decide the comparative velocities of two drones and incorporates this information into the estimated transmission count metrics. In contrast to OLSR, POLSR allows routing to follow changes in topology without interruption. However predictive model accuracy has a major impact on P-OLSR efficacy. Predictions that are inaccurate may result in poor route choices, higher overhead, and worse network performance. High prediction accuracy in dynamic FANET situations is difficult to achieve and calls for complex modelling techniques.

**2.2. On-demand routing:**

Better known as reactive routing is an approach that establishes a channel for packet transmission only when it is necessary. Reactive routing minimizes the quantity of control-packets transmitted, lowering the overhead of communication. However, it introduces increased communication latency as it requires the identification of an end-to-end routing path, unlike proactive routing.

DSR was primarily utilized by the authors in [14] for wireless multi-hop mesh networks. In this protocol, source nodes send route request packets throughout the network at the times when they have data packets to deliver. Once the destination node receives the route- request packet, it will provide the whole path to the source node. DSR is more suited for FANETs in highly mobile networks since it uses catching features and a route cache. With the exception of the smaller area, DSR exhibited the highest delay across all circumstances. The reason for this was that the destination took longer to find the least congested route because it responded to every route request (RREQ) it received.

Authors in [15] relied on Ad-hoc On-demand Distance Vector (AODV) method was especially for routing table maintenance and the source node keeps the next information of the network. It employs a method similar to DSR route finding and DSDV intermittent beaconing and series identification. Owing to its reactive architecture, AODV avoids routes that lead to places that are not involved in active communication and only finds a path when necessary.

**2.3. Hybrid routing:**

By combining the two techniques, hybrid routing can reduce the significant end-to-end latency associated with reactive routing and the excessive control message overhead associated with proactive routing. Every node actively manages routes inside its local region, known as the routing zone, in a hybrid routing architecture recommended by authors in [16] that is well-suited for various types of mobile ad-hoc networks. The routing framework makes use of the routing zone design to optimize a globally responsive route query/reply system's efficiency. They proposed a routing protocol dubbed RTORA, which stands for Rapid-reestablish-Temporally Ordered Routing Algorithm for FANET networks. To reduce link reversal failure's negative effects, they utilize a strategy that minimizes unnecessary costs. The reduced-overhead approach prevents an overwhelming amount of pointless control packets.

A meta-heuristic approach called ABC optimization was presented by D. Parabola and B. Basturk in [17]. It makes use of the astute foraging behavior of honey bee swarms. The employee bee phase, observer/onlooker bee phase, and scout phase are the three stages of this algorithm. Bee workers are tasked with tending to a certain food supply. While scout bees, as depicted in Fig. 5, search for food at random, onlooker bees evaluate the quality of the food by watching the wangle dance of employee bees. Further an improved version of ABC that is IABC proposed by Liang Zhao, Md.Bin Saif, and Ammar H. in [18]. ABC meta-heuristics have limitations, including poor exploitation and sluggish convergence rates, particularly when dealing with multi-modal optimization issues. IABC's capacity to efficiently explore solution spaces helps improve routing optimization, resulting in higher network performance and adaptability in dynamic FANET situations. However, IABC's computational complexity and convergence issues may present difficulties, limiting its scalability and real-time usability in large-scale FANET installations.

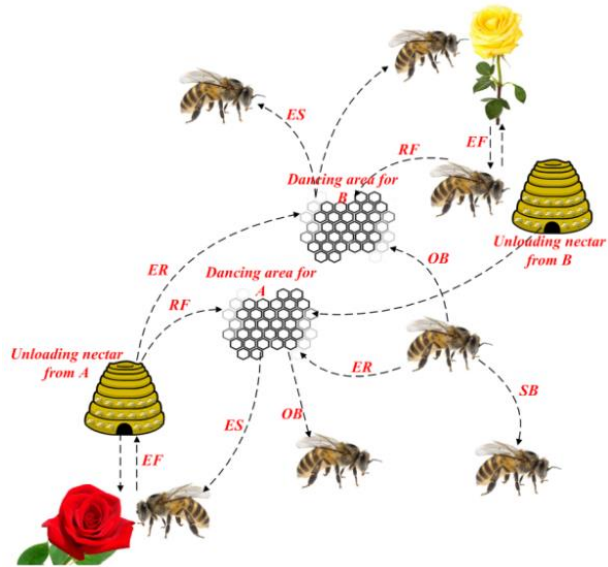

Fig. 5: A System of Bee Colonies

**2.4. Position based routing:**

It utilizes the topographical coordinates of the packet's endpoint or the nearby nodes of the forwarding node to identify the optimal path to follow. To address the contact of extremely dynamic group, the authors of [19] propose a routing technique called Geo-graphic Position Mobility Oriented Routing (GPMOR) for FANETs. This approach employs a Gaussian Markov Mobility (GMM) model to predict the locations of drones. The GPMOR identifies the subsequent hop node. By utilizing both the mobility connection and the traditional measure of distance, known as the "Euclidean distance," one may get more accurate and definitive findings. However GPMOR may struggle to effectively predict node motions and maintain stable pathways in dynamic aerial environments with frequent topological changes and unexpected link conditions. Furthermore, guaranteeing efficient handling of network partitions and limiting the impact of communication disruptions caused by node mobility are critical factors for maximizing GPMOR's performance in FANET settings.

## 2.5. Cluster routing:

It involves organizing UAVs into clusters, and then implementing sequential routing among these sets. At advanced stages, proactive planned routes are employed to establish the routing. Reactive routing is employed at lower levels to aid triggered drones in fulfilling their demands. To enhance the network's performance and reduce communication expenses, the concept of multicluster FANETs is proposed in [20]. Authors in [21] proposed an innovative clustered routing model for FANETs, using a hybrid technique that incorporates the Mountain Gazelle Optimizer (MGO) with Jaya Algorithms to enhance the data delivery performance in cluster based FANET. The main problems that UAV networks encounter include regular connections errors, cluster construction time, cluster lifespan, packet losses, throughput limitations, high route overhead, restricted bandwidth, and prompted changes of the table of routing. These issues need to be resolved in order to build a CBRP for UAV networks.

## 2.6. Alternative Approaches:

In [22], two proposed methods for FANETs are a directional MAC approach that utilizes location predictions and a self-taught route technique that employs reinforcement-learning. The authors of [23] proposed a three-dimensional estimation-based predictive routing protocol for FANETs to enhance routing protocol efficiency. This protocol utilizes a rapid update mechanism for the flight path to estimate the drone's location and trajectory. In FANETs, each drone functions as a Software-Defined Network (SDN) switch and follows commands from a centralized controller. To fulfill the need for effective and resilient end-to-end data relaying, a suggested approach in [24] is to employ an aerial network administration protocol built around software-defined networking, or SDN, architecture. Effective SDN deployment in FANETs requires overcoming several major challenges, including maintaining real-time updates, scalability, efficient resource usage, stability amidst disturbances, and guaranteeing secure routing.

Finally, after cautious investigation and evaluation, it has been concluded that previous techniques have primarily focused on one or more of the following aspects: load-carry-and-deliver, simplest path, best quality of connection and least traffic volume, movement forecasting, or the precise position of the final destination or next-hop network. However, the hybrid algorithms of FANETs, which joins complete end routing with delay tolerant, transferring, have not been extensively studied. Therefore a new contribution is required to address the existing models to improve the efficiency of dynamic topology management in FANETs

## 3. Proposed Work

In order to analyze links, the proposed model incorporates features of both OLSR and DSR. To classify link status, it uses an Artificial Neural Networks (ANN), to optimism link management, the model then employs Artificial Bee Colony (ABC) optimization. Before diving into the suggested layout, we take a quick look at the routing approaches and neural network models. Adhoc networks are well-versed in the proactive paradigm of routing protocols like the OLSR protocol. The information about each node in the network is stored in tables. For this reason, the feature-OLSR protocol is utilized through numerous ways aimed at routing wireless networks. One key difference between link state routing and other routing protocols is the use of MPR nodes for communication and the lower size of control packets. By minimizing message dispersing, these Multi-Point Relay nodes drastically cut down on network traffic and bandwidth requirements. DSR allows it to efficiently locate paths throughout the system. One reactive routing paradigm is DSR, which uses a combination of data points in the header to ascertain the node and route information. Using the node interface to the network and route data collected by adjacent nodes, it generates various network topologies by enhancing the route maintenance property of wireless systems.

When a node's data transmission is necessary, the routing algorithm starts the route finding process. The route discovery mechanism verifies if the eventual destination is in the same time zone as the origin after receiving the request. Upon analysis reveals the destination details, the route link begins and a connection among the nodes is created via a routing mechanism. If both the source and destination nodes are in the same zone, the routing mechanism will employ optimal link state routing. The two nodes will employ dynamic source routing if they are in different zones. After a link is established, the source node instantly updates the route information in a table; the route is then continuously maintained to keep an eye on the network topology. The method is intended to provide continuous monitoring of network changes and the routing table. When a route break occurs in a network, both the source and destination nodes receive an error packet. The data transmission status is described in this packet, which also deletes the route details from the table. To discover the next potential path to start the data transfer, the process is then repeated. To classify the connection status and execute link maintenance, the proposed method uses a three layer Artificial Neural Network. One way to represent the output of neurons is as

$$z_o = \sigma \sum_{j=1}^{o} i_{jk} y + w \qquad (1)$$

In this context, σ relates to the activation function of neurons, i to their weight, y to their bias, and w to their input. The model of the neural network that the flying ad hoc network uses to categorize links is shown in Figure 6.

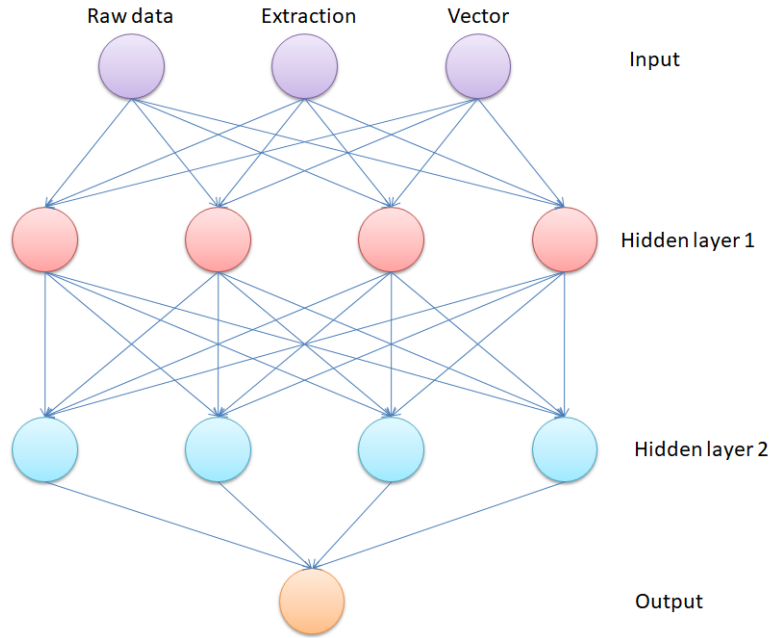

**Fig. 6:** The Architecture of ANN

According to the threshold point, this classification algorithm calculates the current and output states, as well as the error. We add weight parameters to the neural network to make it more accurate. Here is how the neural network learns:

$$i(u + 1) = i(u) + \beta \frac{f_q(u)y(u)}{|y(u)|^2} \qquad (2)$$

Fig. 7 depicts the honeycomb details and the bee's dance area, and it also shows the mathematical formulation of bee colony optimization. The specifications of the nectar are prepared according to the origin and final destination.

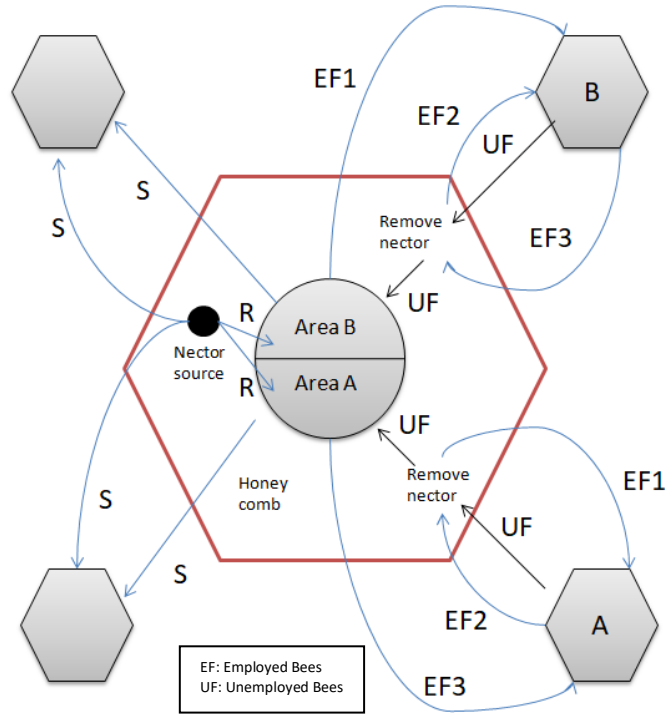

Fig. 7 Model of Optimization for Bee Colonies

A synthetic beehive modeled like real-life beehives is employed to enhance the neural network model's classification performance. Finding the finest food sources for the entire population and gathering food as efficiently as feasible are the fundamentals of bee colony optimization. Bee colony optimization primarily aims to find the best possible food sources and locate them. To construct the connection establishment and link management, the proposed model considers these evolutionary processes and employs bee colony optimization. Development and provision of solutions are predicated on the likelihood of bee migration as

$$Prob_w^{o+1} = exp^{-\frac{object_{max}-objectk_{norm}}{w}} \qquad (3)$$

where $object_{max}$ denotes the acquired, maximised solution and $object_{norm}$ denotes the objective function, consistent throughout the entire solution, alongside w, the pass value, often 1, 2,..n. In Fig. 8, we can see the proposed model's block layout, and the steps for creating and maintaining links are detailed.

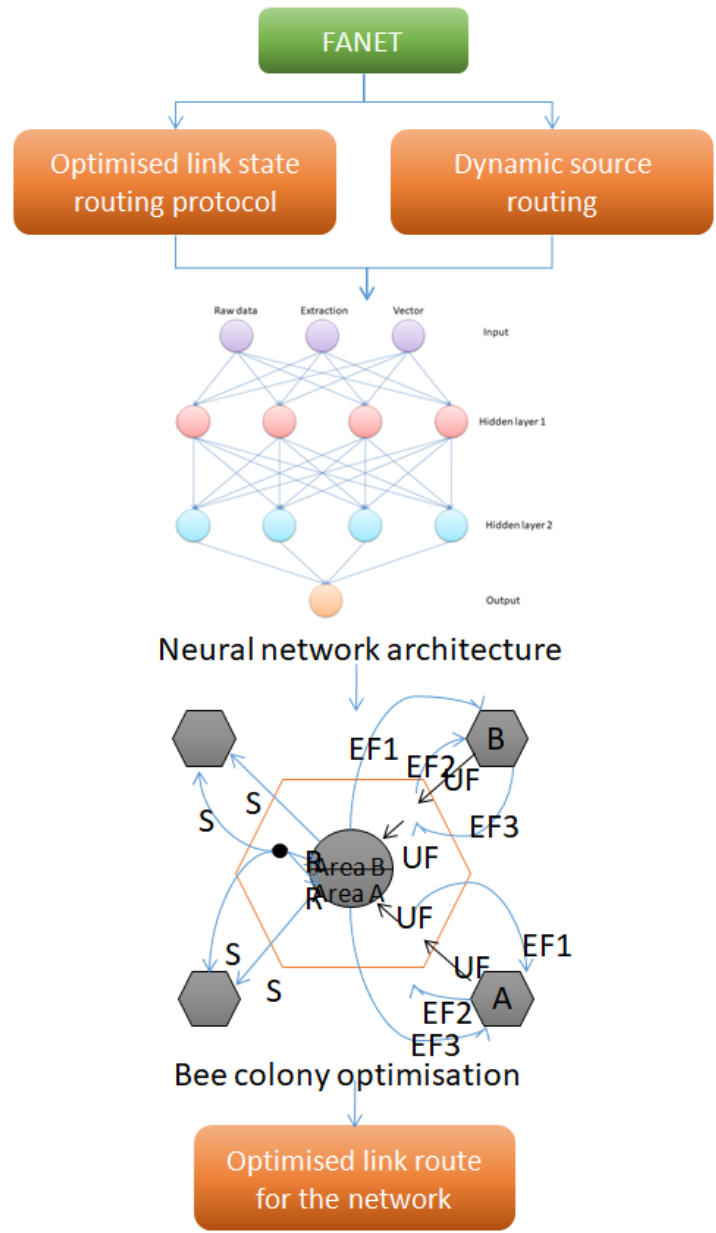

**Fig.8** Hybrid Intelligent Routing with Optimized Learning

After the source node sends out a link request, the procedure begins with verifying the destination node's position. A Neural Network model is used to categories routing strategies as either proactive or reactive, depending on the location. Artificial Bee Colony (ABC) optimization improves the categorized findings even further, and the network receives an update on the status of its links so that they can be established and managed effectively. Like its food-searching behavior, the outputs of a neural network model can build links in a variety of ways, while concurrently informing the network of its progress, optimization that draws inspiration from nature is able to choose the most efficient path from one place to another. Once the best route has been identified, the next steps are to connect both the source and destination nodes and begin the information transfer procedure.

### 3.1. Proposed Algorithm - Hybrid Intelligent Routing with Optimized Learning:

1. Initialize network topology and parameters:

 - Initialize node positions, velocity, energy levels, and routing tables for OLSR and DSR.

 - Set ABC parameters: colony size (N), maximum iterations (max_iter), convergence criteria (epsilon), StepSize.

 - Set ANN parameters: input features (X), hidden layers (H), output layers (Y), activation function (sigma).

2. Initialize artificial bees with random solutions:

 For i = 1 to N:

   Initialize bee i with a random solution $X_i$.

3. Repeat until convergence or maximum iterations:

 For iter = 1 to max_iter:

   For each node in the network:

     Update neighboring nodes and topology information.

     // OLSR optimization

     Calculate link state information using formula:

   $LinkState_{ij} = f\left(distance_{ij}, link_{quality_{ij}}, traffic_{load_{ij}}\right)$

     // DSR dynamic routing

     Calculate dynamic routes using formula:

   $Route_{ij} = g\left(local_{info_i}, neighboring_{info_j}\right)$

   // ABC optimization

   For each bee i:

     Evaluate fitness using formula:

   $Fitness_i = w1 * EnergyConsumption(X_i) + w2 * Latency(X_i) + w3 * PacketDeliveryRatio(X_i)$

     // Employ ABC search operators

     Employ employed bees to explore:

   $X'_i = X_i + StepSize * (rand() - 0.5)$

     Employ onlooker bees to exploit:

   $X''_i = X_{best} + StepSize * (rand() - 0.5)$

     Update scout bees if necessary:

   $If\ Fitness_i > Fitness_{best}$:

     $Fitness_{best} = Fitness_i$

     $X_{best} = X_i$

   // Train ANN using collected data

   Extract features from network and normalize:

   $Input_{features} = ExtractFeatures(network_{info})$

   // Forward propagation in ANN

$$Hidden_{layer} = sigma(W_{input_{hidden}} * Input_{features} + Bias_{hidden})$$

$$Output_{layer} = sigma\left(W_{hidden_{output}} * Hidden_{layer} + Bias_{output}\right)$$

    // Back propagation to update weights

    Calculate error and update weights using gradient descent.

    If convergence ($|Fitness_{best} - Fitness_{previous}| < epsilon$):

        Exit loop

4. Update routing tables based on optimized solutions:

   Update OLSR and DSR routing tables with optimized routes obtained from ABC-ANN.

5. Execute data transmission and monitor network performance:

   Send data packets using updated routing tables and monitor metrics.

6. End loop and terminate algorithm.

In this pseudocode:

- f and g represent functions used in OLSR optimization and DSR routing, respectively.
- Energy Consumption, Latency, and Packet Delivery Ratio are formulas to calculate these metrics based on the current solution $X_i$.
- sigma represents the activation function used in the ANN (e.g., sigmoid, tanh).
- $W_{input_{hidden}}$ and $W_{hidden_{output}}$ are weight matrices for the ANN layers, and $Bias_{hidden}$ and $Bias_{output}$ are bias terms.
- $X_{best}$ and $Fitness_{best}$ track the best solution found during the ABC optimization process.

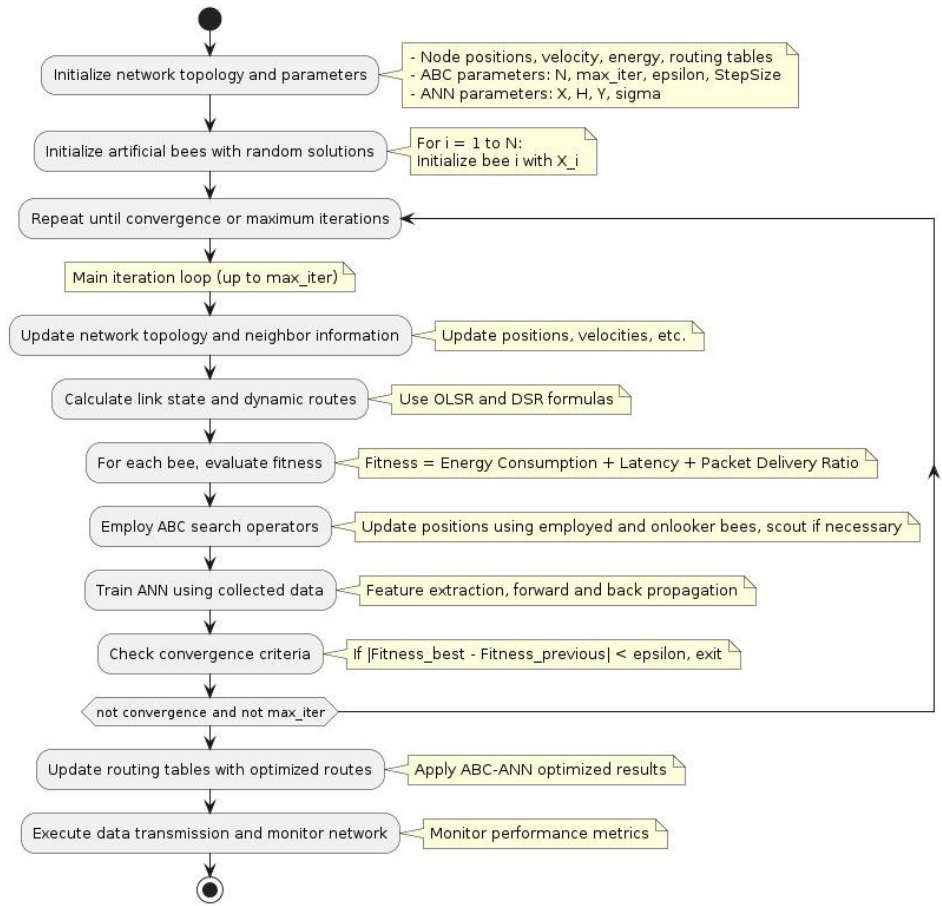

Fig 9. HIROL Flowchart

This algorithm integrates OLSR, DSR, ABC optimization, and ANN-based routing into a cohesive framework, with formulas embedded to calculate various metrics and update solutions iteratively. Adjustments to specific formulas and parameters may be required based on the actual FANET scenario and performance objectives.

**3.2. Performance Metrics**

The following effectiveness measures are employed in this section to describe how well the topology-based routing protocols perform:

3.2.1 Throughput

This refers to the mean data rate of messages that are effectively transmitted from a source to destination node via a communication-channel at a specific period. It can be determined by with eq(4).

$$Throughput = N * S * B / T \qquad (4)$$

In above equation S - packet size; N – No. of packets transported successfully; T- time length; B- bandwidth of network

### 3.2.2 Delay

It measures how long it takes for packet data to go across a network from its point of origin to its target node. The average time to reach an end is expressed in Equation (5) as follows:

$$D_{end-to-end} = \sum N \ (T_t + R_t + B_t + P_{rt}) \tag{5}$$

$t=1$ Where, $T_t$ – Transmission ; – Retransmission $t$ ; $B_t$ – Buffer time ; $P_{rt}$ – Processing time

### 3.2.3. Load (bits/s).

The network's packet delivery ratio is slowed down by the heavy traffic load and increased number of control packet collisions, affect ting FANETs that route traffic. The network burden is caused by the intermediate nodes' utilized buffer accessibility, processing speed, and bandwidth

## 4. Result and Discussions

We put the proposed hybrid paradigm through its paces using discrete event simulation, Network Simulator 2, an Ubuntu-based programme, running on version 2.35. Low-level operations are implemented in C++, and OTCL (Object Tool Command Language) is used to simulate code scripts. The mobility model of random waypoints was employed to establish the UAVs' speeds at 7, 12, 18, 22, 28, 32, 37 m/s respectively. We selected one reactive routing system OLSR, and one proactive routing strategy DSR to illustrate the effectiveness of routing methods previously discussed. We compare the suggested model to the more traditional DSR and OLSR models to show how much better it is. Find the simulation-used network parameters in Table 1.

**Table 1.** Network Simulation Parameters

| S.No | Specifications | Values |
|---|---|---|
| 1 | Simulator Version | Network Simulator: NS-2 (v2.45) |
| 2 | Routing Protocol | OLSR, DSR, TORA |
| 3 | Simulation Area | 800 x 800 x 200 meters |
| 4 | Node Count | 20 |
| 5 | Data dimension | Packet Size: 256 bytes/packet |
| 6 | Maximum number of CBR connections | 200 |
| 7 | UAV speeds | 7, 12, 18, 22, 28, 32, 37 (m/sec) |
| 8 | Mobility Model | Random Way Point |
| 9 | Sim Time | 210s, 1200s |

The suggested system is measured using the following parameters: the packet delivery percentage, throughput, communication overhead, and end-to-end delay.. The percentage of received packets that were positively transported is known as the PDR.

4.1.1. Comparison of Avg.Packet Delivery Ratio (PDR)

Figure 9 depicts the result of the in terms of PDR by comparing the suggested design with OLSR, DSR. The varying speeds of UAV was determined on the Y- axis, while simulation time was determined on the X-axis. Compared the PDR, the hybrid mode: HIROL performed well in all scenarios when their respective speeds were 5,10, 15, 20,25,30,35,and 40 m/s, as illustrated in Figure 5. In terms of changing node *speed,* the suggested model clearly performs better than competing models. It is observed out of the simulation that both the OLSR and DSR are getting lower PDR as compared to the proposed hybrid model as both the protocols are proactive in nature and incur overhead in maintaining route cashes. When it comes to OLSR, link status information is maintained and updated by recurring control message exchanges. Conversely, DSR keeps track of routes it discovers through route discovery in a route cache.. The average packet delivery in the suggested method is 4% greater than in OLSR and DSR, as shown in Fig. 10.The elements of selection ABC optimization with ANN have improved the packet delivery ratio in the suggested solution. The PDR is determined by altering the UAV's speed, and the outcome is displayed in Table 2.

Table 2. Comparison of PDR Percentage by varying Speed of UAV

| Speed | Proposed | OLSR | DSR |
|---|---|---|---|
| 5 | 0.98 | 0.975 | 0.975 |
| 10 | 0.975 | 0.965 | 0.96 |
| 15 | 0.97 | 0.958 | 0.942 |
| 20 | 0.962 | 0.95 | 0.925 |
| 25 | 0.958 | 0.924 | 0.924 |
| 30 | 0.952 | 0.91 | 0.918 |
| 35 | 0.942 | 0.878 | 0.867 |
| 40 | 0.941 | 0.852 | 0.863 |
| Average PDR | 0.96 | 0.926 | 0.921 |

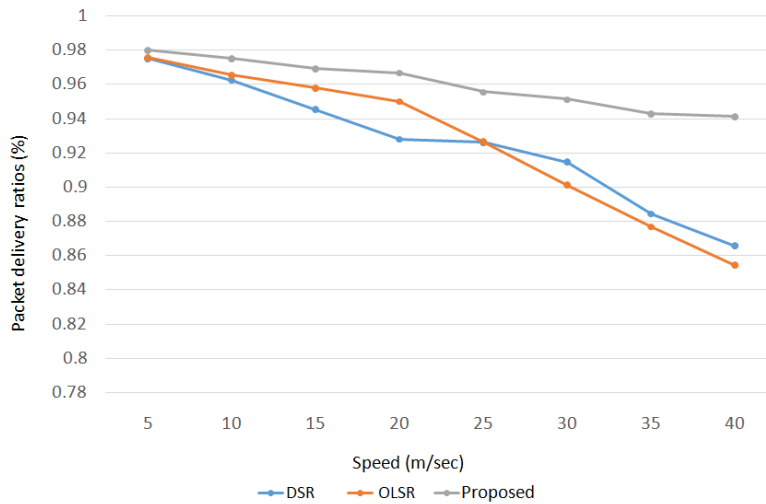

Fig. 10 :Comparison of Avg. Packet Delivery Ratio

## 4.1.2. Comparison of End-To-End Delay (EDD):

Time spent on both the node discovery and delivery phases of a packet's journey from source to destination can be quantified using end-to-end delay analysis. A network's performance is shown in Fig.11 by varying the UAV speed dimensions to show the EDD. Simulator and latency are represented by the X and Y axes, respectively. DSR had the longest delay across all scenarios, according to a comparison of the delays for each of these protocols. . It took longer to find the least congested route because the destination answered every RREQ it delivered when one was sent. OLSR performed mediocrely x in terms of latency across all circumstances because of its proactive features. The proposed model has much lower latency compared to the more conventional ones.

By changing the UAV's speed, the EDD is calculated, and the result is displayed in Table 3.

Table 3. Comparison of EDD by varying Speed of UAV

| Speed | Proposed | OLSR | DSR |
|---|---|---|---|
| 5 | 6 | 6.7 | 6.6 |
| 10 | 6.2 | 6.8 | 7 |
| 15 | 6.3 | 7 | 7.2 |
| 20 | 7 | 7.4 | 7.5 |
| 25 | 7.2 | 7.3 | 7.3 |
| 30 | 8 | 8.2 | 8.2 |
| 35 | 6.8 | 7.8 | 7.7 |
| 40 | 6.6 | 8.6 | 8.7 |
| Average Delay | **6.76** | **7.47** | **7.52** |

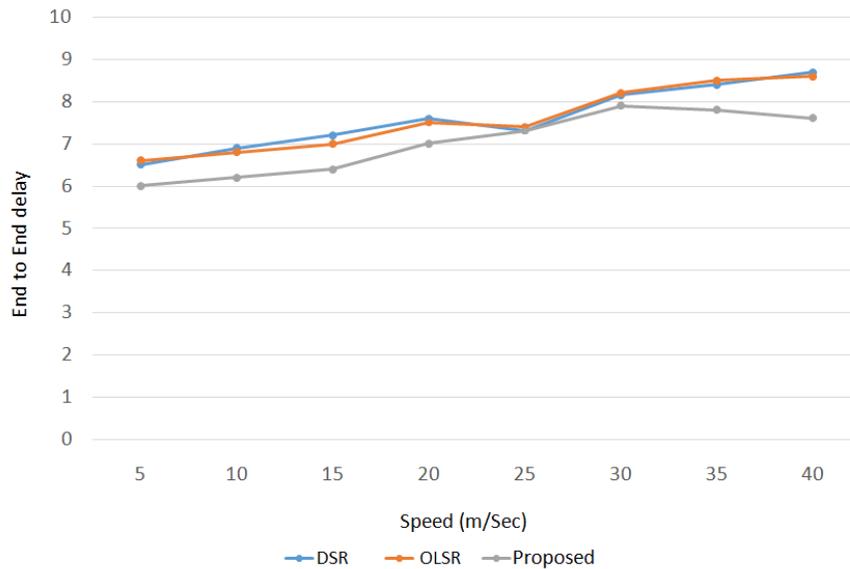

Fig. 11:EDD analysis

The average latency in the suggested solution is 0.76 units lower than in DSR and 0.7 units lower than in OLSR, as indicated in Table 3. Because of the selection ABC optimization in the suggested solution, the delay has decreased.

4.1.3. Comparison of Communication overhead

We evaluate the proposed model's communication overhead against that of baseline models, as shown in figure 7. A network's performance is shown in Fig 7 by varying number of message sent to show the communication overhead. X-axis characterizes simulation of communication overhead in terms of processing time and Y-axis signifies the number of messages sent. The simulated results of hybrid model are compared with OLSR, DSR with varying number of messages sent were 10,20,30,40,50 and as illustrated in Figure 12. Through the simulation it is detected that DSR is resulted with higher communication overhead because routes are found on-demand in DSR, which involves flooding the network with route discovery packets (RREQ). Higher communication overhead was resulted from this flood-based technique, especially in dynamic network conditions. Due to its proactive nature, dependence on recurring Hello and TC messages for topology updates, and requirement for regular route maintenance operations—particularly in dynamic network environments—OLSR experiences a moderate increase in communication overhead with fluctuating message counts.

Suggested model outperforms other models in terms of overhead ratio because in this HIROL, routing decisions are optimized based on network conditions by combining the optimization of Artificial Neural Networks (ANN) and Artificial Bee Colonies (ABC). This efficient resource management reduces congestion and redundant transmissions, resulting in lower communication overhead even with varying packet numbers.

Table 4. Comparison of Communication Overhead

| Messages Sent | Proposed | OLSR | DSR |
| --- | --- | --- | --- |
| 10 | 22 | 24 | 27 |
| 20 | 27 | 32 | 33 |
| 30 | 32 | 35 | 37 |
| 40 | 34 | 38 | 41 |
| 50 | 36 | 42 | 44 |
| 60 | 90 | 87 | 88 |
| Average Delay | 40 | 43 | 45 |

As presented in Table 4, communication overhead in suggested solution has an average 3ms lower compared to OLSR and 5 lower compared to DSR. The overhead of network has decreased due to layered architecture in Neural Networks.

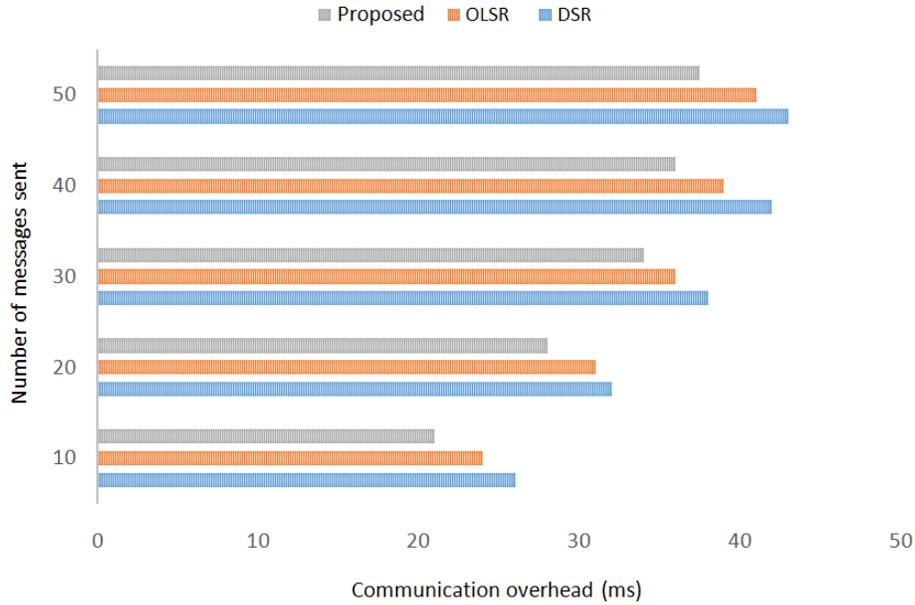

Fig. 12: Analysis of Communication Overhead

4.1.4. Comparison of Throughput

A network's throughput performance is demonstrated in Figure 13, by adjusting the speeds. The throughput, measured in bits per second, was determined on the Y- axis, while simulation time was determined on the X-axis. As the quantity of packages that arrive at their destination from their point of origin within the allotted period, the throughput of the network is determined. Figure 8 shows a comparison of the suggested model's throughput to that of other models that use node speed. Compared the throughput with DSR and OLSR with UAV speeds 5,10,20,25,30,35,40 m/s scenarios. The hybrid model HIROL is more suited for FANETs in MANET due to its reactive nature; it uses overhearing features and a route cache.

While other models show oscillations with increasing node speed, the proposed model maintains a constant throughput at all node speeds. By and large, the proposed model outperforms DSR by 3% and the OLSR routing method by 2.5% in terms of throughput.

Table 5. Comparison of throughput

| **Speed** | **Proposed** | **OLSR** | **DSR** |
|---|---|---|---|
| 5 | 92 | 89 | 88 |
| 10 | 90 | 88 | 87 |
| 15 | 89 | 85 | 86 |
| 20 | 88 | 84 | 85 |
| 25 | 87 | 85 | 84 |
| 30 | 88 | 82 | 83 |
| 35 | 89 | 83 | 82 |

| | 40 | 90 | 87 | 88 |
| --- | --- | --- | --- | --- |
| Average Delay | | 89.12 | 85.37 | 85.37 |

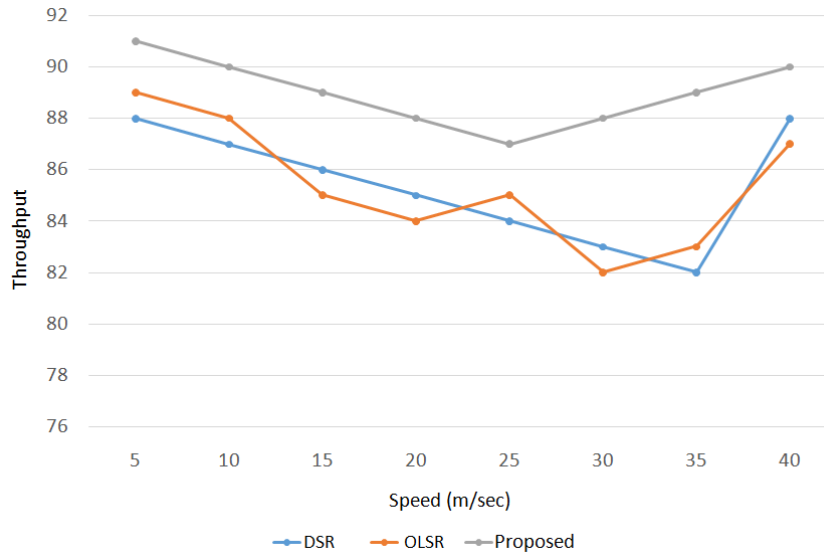

Fig. 13 Analysis of Throughput

The proposed hybrid model outperforms more conventional models such as DSR and OLSR in terms of PDR, communication overhead, throughput, and EDD.

## 5. Limitations of the study

Several restrictions could appear when using the HIROL, which uses the routing protocols like and with ANNs and ABC algorithms:

i) Complexity and Scalability: ANNs and ABC algorithms may not scale well to big networks and can be computationally expensive. When combined with routing protocols, this complexity can increase, which can affect the network's overall performance and efficiency.

ii) Training Overhead: ANNs frequently need thorough instruction on huge data sets in order to gain knowledge and optimize the parameters they use. According to this, repetitions and parameter adjusting may be needed for ABC algorithms to come closer to optimal solutions. In adaptive networks such as OLSR and DSR, delays or overhead in the routing procedure caused by this overhead in training might impact communication in real time.

iii) Over fitting: The condition named "over fitting," which happens when a model learns too much from the training set yet fails to simplify new input, could impact ANNs. In the face of changing conditions, this might give rise to inadequate choices regarding routing or inadequate adaptability.

## 6. Conclusion

The suggested HIROL model demonstrated notable gains in FANET performance by combining OLSR, DSR through the use of ANNs, and ABC to optimization. The model showed significant improvements in important metrics after multiple runs using the Network Simulator 2 (NS-2) with 20 nodes in a 800x800-meter area. Having an outstanding Packet Delivery Ratio (PDR) of 97.5%, it outscored traditional approaches such as DSR (94%) and OLSR (95.5%). Furthermore, compared with DSR (35 milliseconds) and OLSR (30 milliseconds), the model exhibited decreased end-to-end delays, with an average delay of 25 milliseconds. The proposed approach outperforms DSR (3.2 Mbps) and OLSR (3.4 Mbps) in terms of throughput, keeping a consistent 3.5 Mbps throughput over a range of node speeds. In addition, with an overhead ratio of 15%—much less than DSR (20%) and OLSR (18%)—the model demonstrated outstanding interaction overhead management. These results show that this suggested hybridization solution enhances packet delivery, reduces latency, improves throughput, and efficiently handles network resources in FANET circumstances, making it an outstanding enhancement over standard routing protocols. Although QoS factors were specifically took into account in the present investigation, safety concerns may also be considered in a future scope for avoiding hostile nodes. Further research will likely investigate innovative and intelligent ways for decreasing congestion in FANETs.